\documentclass[a4 paper, 9pt,reqno]{amsart}

\usepackage{amsmath,amssymb}
\usepackage{latexsym}
\usepackage{epsfig}
\usepackage{cite}
\usepackage[english]{babel}
\usepackage{booktabs}
\usepackage{graphicx}
\usepackage{tikz}
\usepackage{pgfplots}
\usepackage{eurosym}
\usepackage[inner=4cm,outer=2cm]{geometry} 
\usepackage{amsthm}
\usepackage{MnSymbol}

\usepackage{pifont}
\usepackage{verbatim} 
\usepackage{enumerate}
\usepackage{geometry}
\usepackage{color}
\numberwithin{equation}{section}
\title{Super throats with non trivial scalars}

\author{D. Astesiano}
\address{Universit\`a  dell'Insubria, Dipartimento di Scienza ed Alta Tecnologia, via Valleggio 11, 22100, Como, Italy, and INFN via Celoria 16, 20133, Milano, Italy}
\email{dastesiano@uninsubria.it}

\author{S. L. Cacciatori}
\address{Universit\`a  dell'Insubria, Dipartimento di Scienza ed Alta Tecnologia, via Valleggio 11, 22100, Como, Italy, and INFN via Celoria 16, 20133, Milano, Italy}
\email{sergio.cacciatori@uninsubria.it}

\begin{document}

\begin{abstract}
{We find new BPS solutions in $N=2$ $D=4$ Fayet-Iliopoulos gauged supergravity with STU prepotential. These are stationary solutions carrying a Kerr-Newman throat spacetime geometry and are everywhere regular. One of the three scalar vector 
fields is non constant. Moreover, they carry non vanishing magnetic fluxes and dipolar electric fields.}
\end{abstract}

\maketitle

\section{Introduction}
The relevance of black holes in AdS spaces is known for several reasons. One is of course AdS/CFT and its several applications, as for example to condensed matter physics, see \cite{Hartnoll}, Fermi liquids physics,
\cite{Iizuka}, and superconductivity, \cite{Hartnoll2}. The coupling to electromagnetic charges and scalar field is crucial in these constructions, at least for looking for realistic physical models, so, considering backgrounds in gauged supergravity including 
abelian gauge fields and coupled to scalar matter is quite natural. 
BPS solutions provide examples where supersymmetric conformal field theories are defined on curved backgrounds, the conformal boundaries. Also non BPS and even non extremal solutions are of interest for holographic construction 
related, for example, to finite temperature condensed matter systems. Another issue is Kerr/CFT correspondence, which allows to provide a microscopic description and computation of the black hole entropy,
\cite{Guica}, \cite{Benini}. 
A related phenomenon is the attractor mechanism that for AdS black holes is quite different than the quite well understood case of asymptotically flat black holes, \cite{Bellucci}, \cite{Cacciatori2}, \cite{Kachru}.
In the recent years there have been progresses in finding BPS, non BPS and thermal black holes solutions in $N=2$ gauged supergravity in four dimensions, coupled with matter, see for example 
\cite{Hristov}, \cite{Gnecchi}, \cite{Cacciatori}
\cite{Daniele}, \cite{Klemm}, \cite{Chow-1}, \cite{Chow-2}, \cite{Chow-3}, \cite{Chong} and \cite{Hristov-1}. 
A way of finding BPS solutions is to directly face the equations deduced  in \cite{Cacciatori}, where all timelike BPS equations are classified for $N=2$, $D=4$ gauged supergravity coupled to an arbitrary number of abelian vector multiplets.
This is the strategy used here, even though a more general setting could be to follow the strategy of \cite{Gnecchi}, where a very general structure for (non necessarily BPS) black hole solutions has been individuated and proposed as a possible
general structure of the black hole solutions.\\
Black hole solutions, however, are not the only interesting ones. Indeed, it is well known the role of Bertotti-Robinson solutions in supergravity, in particular, in relation to the attractor mechanism, \cite{Ferrara:1997yr}. They typically represent the near 
horizon geometries of extremal static black holes. In the stationary case, the near horizon geometry may become more involved but, as shown in \cite{Bardeen}, it has several characteristics in common with the Bertotti-Robinson geometry. Indeed, these 
kind of geometries have been investigated also in different dimensions, see for example \cite{Clement:2001gia}, \cite{Bouchareb:2014fxa}, \cite{Clement:2001ny}, and \cite{Kunduri:2007vf}.\\
Here we will present a different class of BPS solutions, which are everywhere regular solutions with non constant scalar fields. More precisely, we consider the Fayet-Iliopoulos 
gauged supergravity coupled to three vector multiplets in the STU model with prepotential $F(X^0,X^1,X^2,X^3)=-2i\sqrt{X^0X^1X^2X^3}$. Our solutions looks like deformed Bertotti Robinson spacetimes: they are fibrations of $AdS_2$ spaces
over a genus zero Riemann surface with non constant scalar curvature. These are stationary solutions carrying magnetic fluxes. 
There are electric fields also but without fluxes and there are not electric and magnetic charges.
In particular, the electric fields fluxes 
can be switched off by varying a parameter to zero, related to the imaginary part of the non constant scalar field.
When the electric fields are switched off the $AdS$ fibres become flat Minkowski spacetimes $\mathbb R^{1,1}$.\\
We also find a static solution. Moreover, there are also other similar solutions with the same structure but with the base replaced by non compact spaces. We however did not studied them since they do not admit the action of finite groups allowing for
a compactification or to a reduction of a finite area of the horizon,  and we considered the compact ones more interesting.
In any case, all these solutions, even if not black holes, are in the general form proposed in equation (2.1) of \cite{Gnecchi}.

\section{General conventions and equations of motion}
We will follow the conventions in \cite{Cacciatori2}.
Let us consider $\mathcal N=2, D=4$ gauged supergravity coupled to $n_V$ abelian vector multiplets. Its bosonic content is given by the vierbein $e^a_\mu$, the $U(1)$ gauge vectors $A^I_\mu$, $I=0,\ldots,n_v$ and complex scalar fields $z^\alpha$, 
$\alpha=1,\ldots,n_V$, parameterising
a special K\"ahler manifold, which is the base of a symplectic bundle having covariantly holomorphic sections
\begin{align}
\mathcal V=
\begin{pmatrix}
 X^I\\ F_I
\end{pmatrix}, \qquad\ \mathcal D_{\bar \alpha} \mathcal V =\partial_{\bar \alpha} \mathcal V -\frac 12 (\partial_{\bar \alpha} \mathcal K) \mathcal V=0.
\end{align}
Here $\mathcal D$ is the K\"ahler covariant derivative and $\mathcal K$ the K\"ahler potential, and $\partial_{\bar\alpha}$ is the partial derivative w.r.t. $\bar z^{\bar\alpha}$. These sections are constrained by
\begin{eqnarray}
\langle \mathcal V, \bar {\mathcal V} \rangle=X^I \bar F_I-F_I \bar X^I=i.
\end{eqnarray}
Usually one assumes the existence of a homogeneous function $F$ of degree two, called the prepotential, such that
\begin{align}
 v(z):=e^{-\mathcal K(z,\bar z )/2} \mathcal V= 
\begin{pmatrix}
 Z^I(z) \\ \frac {\partial F(Z)}{\partial Z^I}
\end{pmatrix}.
\end{align}
From this we see that the K\"ahler potential can be computed as
\begin{align}
 e^{-\mathcal K(z,\bar z )}=-i\langle v,\bar v \rangle.
\end{align}
The scalars $z^\alpha$ are coupled to the gauge fields via the period matrix $\mathcal N_{IJ}$ defined by
\begin{align}
 F_I=\mathcal N_{IJ} X^J, \qquad\ \mathcal D_{\bar \alpha} \bar F_I=\mathcal N_{IJ} \mathcal D_{\bar \alpha} \bar X^J.
\end{align}
The bosonic part $\mathcal L_b$ of the lagrangian density is 
\begin{align}
 \frac {\mathcal L_b}{e}=&\frac R{2} + \frac 14 (Im \mathcal N)_{IJ} G^I_{\mu\nu} G^{J\mu\nu} - \frac 1{8e} (Re \mathcal N)_{IJ} \varepsilon^{\mu\nu\rho\sigma} G^I_{\mu\nu}G^J_{\rho\sigma}\cr
 &-g_{\alpha\bar \beta}\partial_\mu z^\alpha \partial^\mu 
 \bar z^{\bar \beta} -V,
\end{align}
where $G^I_{\mu\nu}=\partial_\mu A^I_\nu-\partial_\nu A^I_\mu$ are the field strengths and $V$ is the scalar potential
\begin{align}
 V=-2g_I g_J \left[ (Im \mathcal N)^{-1|IJ}+8 \bar X^I X^J \right],
\end{align}
with $g_I=g\xi_I$ constant.
In \cite{Cacciatori} it has been shown that if one looks for supersymmetric solutions admitting a timelike Killing spinor, then the most general supersymmetric background can be expressed in coordinates $t,z, w,\bar w$ as
\begin{eqnarray}
ds^2=-4|b|^2 (dt+\sigma)^2+|b|^{-2} (dz^2+e^{2\Phi}dwd\bar w)
\end{eqnarray}
where $b(z,w,\bar w)$ is a complex function, $\Phi(z,w,\bar w)$ a real function and $\sigma=\sigma_w dw+\sigma_{\bar w} d\bar w+\sigma_z dz$ a one form.\\
After setting
\begin{align}
 \mathcal I=Im(\mathcal V/\bar b),
\end{align}
the fields must satisfy the following system of coupled nonlinear equations
\begin{align}
d\sigma +2 \star^{(3)} \langle \mathcal I , d\mathcal I \rangle -\frac {i}{|b|^2}g_I \left( \frac {\bar X^I}{b} +\frac {X^I}{\bar b} \right)  e^{2\Phi} dw \wedge d\bar w &=0, \label{1} \\
 \partial_z \Phi+2ig_I \left( \frac {X^I}{\bar b}- \frac {\bar X^I}{b} \right)&=0, \label{2} \\
 4\partial_w \partial_{\bar w} \left( \frac {X^I}{\bar b}- \frac {\bar X^I}{b} \right) +\partial_z \left[e^{2\Phi} \partial_z \left( \frac {X^I}{\bar b}- \frac {\bar X^I}{b} \right)\right]+ & \cr  -2ig_J \partial_z
 \left\{ e^{2\Phi} \left[ |b|^{-2} (Im \mathcal N)^{-1|IJ} +2\left( \frac {X^J}{\bar b}+ \frac {\bar X^J}{b} \right) \left( \frac {X^I}{\bar b}+ \frac {\bar X^I}{b}\right) \right] \right\}&=0, \label{3} \\
 4\partial_w \partial_{\bar w} \left( \frac {F_I}{\bar b}- \frac {\bar F_I}{b} \right) +\partial_z \left[e^{2\Phi} \partial_z \left( \frac {F_I}{\bar b}- \frac {\bar F_I}{b} \right)\right]+ & \cr  -2ig_J \partial_z
 \left\{ e^{2\Phi} \left[ |b|^{-2} Re\mathcal N_{IL} (Im \mathcal N)^{-1|JL} +2\left( \frac {F_I}{\bar b}+ \frac {\bar F_I}{b} \right) \left( \frac {X^J}{\bar b}+ \frac {\bar X^J}{b}\right) \right] \right\}+& \cr
 -8ig_Ie^{2\phi}\left[ \langle \mathcal I , \partial_z \mathcal I \rangle -\frac {g_J}{|b|^2} \left( \frac { X^J}{\bar b} +\frac {\bar X^J}{ b} \right) \right]  &=0, \label{4} \\
 2\partial_w\partial_{\bar w}\Phi -e^{2\Phi} \left[ ig_I \partial_z \left( \frac {X^I}{\bar b} -\frac {\bar X^I}{b} \right) +\frac 2{|b|^2} g_I g_J (Im \mathcal N)^{-1|IJ} \right. & \cr
 \left.+4 \left( \frac {g_I X^I}{\bar b} +\frac {g_I\bar X^I}{b} \right)^2 \right]&=0. \label{5}
\end{align}
Here $\star^{(3)}$ is the Hodge star on the three-dimensional base with metric 
\begin{equation}
ds^2_3 = dz^2 + e^{2\Phi} dw d\bar w. 
\end{equation}
The fieldstrengths corresponding to a given solution $b$, $\Phi$, $\sigma$ and $\mathcal V$, are:
\begin{align}
 &G^I= 2(dt+\sigma)\wedge d(bX^I+\bar b \bar X^I) \cr
 & +|b|^{-2} \left[ \bar X^I(\partial_{\bar w}\bar b+iA_{\bar w} \bar b)+(\mathcal D_\alpha X^I)b\partial_{\bar w} z^\alpha -X^I(\partial_{\bar w}b -iA_{\bar w}b)-(\mathcal D_{\bar \alpha} \bar X^I)\bar b \partial_{\bar w} \bar z^{\bar\alpha} \right]
 dz\wedge d\bar w \cr
 & -|b|^{-2} \left[ \bar X^I(\partial_{w}\bar b+iA_{w} \bar b)+(\mathcal D_\alpha X^I)b\partial_{w} z^\alpha -X^I(\partial_{w}b -iA_{w}b)-(\mathcal D_{\bar \alpha} \bar X^I)\bar b \partial_{w} \bar z^{\bar\alpha} \right]
 dz\wedge d w \cr
 &-\frac 12 |b|^{-2} e^{2\Phi} \left[ \bar X^I(\partial_{z}\bar b+iA_{z} \bar b)+(\mathcal D_\alpha X^I)b\partial_{z} z^\alpha -X^I(\partial_{z}b -iA_{z}b)-(\mathcal D_{\bar \alpha} \bar X^I)\bar b \partial_{z} \bar z^{\bar\alpha} \right. \cr
 &\left. -2ig_J (Im \mathcal N)^{-1|IJ}\right] dw\wedge d\bar w,
 \label{Gauge}
\end{align}
where $A_\mu$ is the gauge field of the Kahler U(1),
\begin{equation}
A_\mu = - \frac{i}{2} ( \partial_\alpha \mathcal K \partial_\mu z^\alpha - \partial_{\bar\alpha}\mathcal K \partial_\mu \bar{z}^{\bar\alpha} ).
\end{equation}
By now we will refer to eq.(\ref{1}) as the rotating equation,  eq.(\ref{2}) as the radial equation, eq.(\ref{3}) as the Bianchi identity, eq.(\ref{4}) as the Maxwell equations, and eq.(\ref{5}) as the metric equation.
\section{$STU$ model}
\noindent
We are interested in looking for stationary solutions with at least one non constant scalar, and characterised by a nontrivial $\sigma$. 
To this end, we consider the $STU$ model with prepotential
\begin{align}
 F=-2i\sqrt{(X^0X^1X^2X^3)}.
\end{align}
The symplectic section can be parametrised in terms of three complex scalar fields $\tau_1$, $\tau_2$ e $\tau_3$ by choosing $Z^0 = 1$, $Z^1= \tau_2 \tau_3$, $Z^2 = \tau_1 \tau_3$, $Z^3 = \tau_1 \tau_2$, so that
\begin{equation}
 v^T= 
\begin{pmatrix}
 1, & \tau_2 \tau_3, & \tau_1 \tau_3, & \tau_1 \tau_2, & -i \tau_1 \tau_2 \tau_3, & -i\tau_1, &  -i \tau_2, & - i\tau_3 
\end{pmatrix}.
\end{equation}
$\tau_\alpha$ and $\bar \tau_{\bar\alpha}$ are the complex coordinates on the scalar manifold.
The K\"ahler potential and the non vanishing components of the metric on the scalar manifold are respectively
\begin{align}
 e^{-\mathcal K} =8 Re\tau^1 Re\tau^2 Re\tau^3  \qquad 
 g_{\alpha\bar\alpha} =g_{\bar\alpha\alpha}=\partial_{\alpha}\partial_{\bar\alpha}  \mathcal K=(\tau_{\alpha}+\bar\tau_{\bar\alpha})^{-2}.
\end{align}
In particular, we notice the relations
\begin{align}
 F_I=\frac {F}{2X^I}, \qquad \mathcal N_{IJ}=\frac {F}{2(X^{I})^2}\delta_{IJ}
\end{align}
between the prepotential and the period matrix. Inspired by \cite{Cacciatori2} and \cite{Klemm}, in order to solve the equations, we propose the ans\"atze 
\begin{align}
    \tau_1&=\sqrt {\frac {g_0g_1}{g_2g_3}}\ \tau(z,w,\bar w), \qquad \tau_2=\sqrt {\frac {g_0g_2}{g_1g_3}}, \qquad \tau_3=\sqrt {\frac {g_0g_3}{g_1g_2}}, \label{tau}\\
    e^{2\Phi}&=h(z)\ell( w,\bar w). \label{Phi}
\end{align}
Compatibly with this choice, we also make the ans\"atze
\begin{align}
    & \frac{\bar X^A}{b} = \frac{\hat{\eta}(z, w, \bar w)}{g_A} \qquad \text{with}\,\, \text{A}=0,1 \,\,
    & \frac{\bar X^B}{b} \equiv \frac{if(z)+\eta(z,w, \bar w)}{g_B} \qquad \text{with}\,\, \text{B}=2,3 \,\,.
\end{align}
Replacing in equations (\ref{2}) and (\ref{5}), we get
\begin{align}
    \frac{\partial_z h}{h} &= -16 f, {\label{10}} \\
    \frac{\partial \bar{\partial} \ln{\ell}}{\ell} &= h \left[-\frac{1}{2} \frac{\partial^2_z \sqrt{h}}{\sqrt{h}} + 32  ( \hat{\eta}^2 + \eta^2 + 4 \hat{\eta} \eta)  \right], {\label{7}}
\end{align}
with the condition $g_B>0$. Moreover, we take $g_A <0$ because of the requests $Re \tau > 0$ and $Im(N_{IJ}) < 0 $.
\newline
Bianchi's equations become, for $I=A$ and for  $I=B$ 
\begin{align}
 \partial_z \left[h(\hat{\eta}^2+ 2 \hat{\eta} \eta)\right]&= 0, \label{8} \\
    \frac{1}{4} \partial_z \left( \sqrt{h} \partial^2_z  \sqrt{h} \right) -16 \partial_z\left[h(\eta^2+ 2 \eta \hat \eta)\right]&=0, \label{9}
\end{align}
respectively.
These are solved by setting\footnote{with the factor 16 included for convenience. This is not the most general solution, but we will discuss the most general case in the Conclusions section}
\begin{align}
       & \eta = \frac 1{16}\frac{ s( w, \bar w)}{g(z)}, \qquad  \hat{\eta} = \frac 1{16} \frac{ \hat s(w , \bar w)}{g(z)}, {\label{s}}
\end{align}
with
\begin{align}
g(z)=\pm \sqrt {h(z)}, \qquad\  h(z)=(\alpha z + \beta )^2 \label{s1},
\end{align}
where $\alpha$ and $\beta$ are arbitrary constants.\footnote{For $\alpha$ different from zero, it is possible to reabsorb both $\alpha$ and $\beta$ in the definition of the variable $z$, but this notation allows us to include the case $\alpha = 0$.}
\newline
The Maxwell equations become 
\begin{gather}
 \frac{1}{\ell}\, \partial \bar\partial s =  \frac{1}{8} s \hat{s}(\hat s+s) \label{MA}, \\
   \frac{1}{\ell}\, \partial \bar\partial \hat s =  \frac{1}{8} s \hat{s}(\hat s+s). 
\end{gather}
It follows immediately that $s-\hat{s}$ is a real armonic function. So, in general we must have
\begin{equation}
    \hat{s}(w,\bar w)= s(w, \bar w) + F(w) + \bar{F}(\bar{w}) \label{F}
\end{equation}
where $F$ is a holomorphic function. Moreover eq.(\ref{7}) reduces to
\begin{align}
\frac{\partial \bar{\partial} \ln{\ell}}{\ell} =   \frac{1}{8} ( \hat{s} + s)^2 + \frac{1}{4} \hat{s} s \label{MB}.
\end{align}
In this way our system of equations is reduced to eq. (\ref{1}), (\ref{MA}), (\ref{F}) and (\ref{MB}).
\section{A Static solution}
A simple static solution is easily obtained putting $\hat{s}= -s = a $, where $a$ is a constant. It is then straightforward to prove that  the general solution of eq.(\ref{MB}) is
\begin{align}
    \ell(x) &= \frac{c}{2 a^2} \left[ 1 - \tanh^2{ \left( \frac{\sqrt{c}}{2} x\right)} \right], 
\end{align}
where $c>0$ is an integration constant.\footnote{We omitted an irrelevant integration constant that would not change the meaning of the solution.} 
\newline
The full solution reads
\begin{align}
    ds^2 &= - \frac{2 G h}{a^2} dt^2 + \frac{2 a^2}{G h} dz^2 + \frac{2 a^2 \ell}{ G} (dx^2 + dy^2), \\
    \tau_1 &= \sqrt{\frac{g_0 g_1}{g_2 g_3}} \left(1+\frac{2i \alpha}{a}\right), \qquad h = (\alpha z + \beta)^2,
\end{align}
where $G= 64 \sqrt{g_0 g_1 g_2 g_3}$. 
\newline
It is convenient to introduce the new variables $\theta$ and $\phi$ such that
\begin{align}
    \sin^2{\theta} &=  1 - \tanh^2{ \left( \frac{\sqrt{c}}{2} x\right)}, \\
    \phi&=\frac{\sqrt{c}}{2} y  .
\end{align}
This way the metric takes the form
\begin{align}
   ds^2 = - \frac{2 G (\alpha z +\beta)^2}{a^2} dt^2 + \frac{2 a^2}{G (\alpha z + \beta)^2} dz^2 + \frac{4}{ G} (d \theta^2 + \sin^2{\theta} \, d\phi^2). 
\end{align}
For $\alpha$ different from zero this is a Bertotti-Robinson space-time, which is a $AdS_2 \times S^2$, with scalar curvature $- \frac{\alpha ^2 G}{a^2}$ for 
$ AdS_2$ and $G/2$ for $S^2$. For $\alpha= 0$, it is an $ \mathbb R^{1,1} \times S^2$. \\
The fields strength are
\begin{align}
   G^A &=-\frac{G \alpha}{8 a g_A} dt \wedge dz + \frac{\sin{\theta}}{ 8 g_A} d\theta \wedge d\phi, \\ 
   G^B &=\frac{G \alpha}{8 a g_B} dt \wedge dz + \frac{\sin{\theta}}{8 g_B} d\theta \wedge d\phi,   
\end{align}
with duals
\begin{align}
   Im \mathcal N_{AJ}  \star G^J &=-4 g_A dt \wedge dz - \frac{16 \alpha g_A \sin{\theta} }{Ga} d\theta \wedge d\phi, \\ 
  Im \mathcal N_{BJ}  \star G^J &=- 4 g_B \frac{a^2}{a^2+4\alpha^2} dt \wedge dz + \frac{16 \alpha g_A \sin{\theta} }{Ga}  \frac{a^2}{a^2+4\alpha^2}  d\theta \wedge d\phi. 
\end{align}
Therefore, the magnetic and electric fluxes are
\begin{align}
& \frac{1}{4 \pi} \int_\sigma G^I = \frac{1}{8 g_I}, \\ 
&  \frac{1}{4 \pi} \int_\sigma ( Re \mathcal N_{IJ} \star G^J -Im \mathcal N_{IJ} \star G^J)=0,
\end{align}
respectively. Since $d \star G^I = 0$, the electric four-current densities $J^I$ vanish. Since the solution is everywhere regular, this means that there are not charges but just free constant fields.\\
This kind of solution is not new and is well understood. In particular, these near horizon metric have been studied in \cite{Bellucci} in general. In particular, our solution solves (2.16)-(2.18) in \cite{Bellucci}, where in our case 
$V\lfloor_{r_H}=\frac G8\left( 1-2\frac {\alpha^2}{a^2} \right)$,  $V_{BH}\lfloor_{r_H}=\frac 2G \left(1+2\frac {\alpha^2}{a^2}\right)$, and $r_H^2=\frac 4G$, $r_A^2=\frac {2a^2}{\alpha^2 G}$. 
It is worth to mention here that in that paper an imprecise statement is made at the end of section 2 and we need to clarify in the present exposition: it is stated that all these solutions, for $R_H\neq R_A$, are not BPS solutions. Of course this 
is not true, as it is well known, and it has to be interpreted as follows. 
These near horizon solutions are 1/2 BPS, so that we have an enhancement of supersymmetry with respect to the expected one. This can be checked by solving explicitly the equations (3.9) in \cite{Cacciatori}, whose solutions
are constant Dirac spinors satisfying the condition $\Gamma^2\Gamma^3 \epsilon=\epsilon$.
Since the near horizon geometry enhances supersymmetry, the full flow solutions have to be less 
supersymmetric and are expected to be 1/4 supersymmetric flows. These are exactly in the general class considered for example in \cite{DallAgata:2010ejj}, even if have not been able to find explicitly a solution restricting exactly to our
configuration as regards, in particular, the electromagnetic fields.

\section{Stationary solutions}
\noindent
Much more interesting solutions to the system (\ref{MA})-(\ref{MB}) can be found just assuming $F$ constant, say $ \hat{s} = s + 2a$ where $a$ is a constant. 
The resulting equations are stationary conditions for the action functional
\begin{equation}
    S[g,s] =  \int dw d\bar w\ \ell(w,\bar w) [ s(w,\bar w) R(w,\bar w) + \psi(s(w,\bar w)) ],
\end{equation}
where $s$ is a scalar function with potential $\psi(s)=2s(s+2a)(s+a)$ and $R$ is the scalar curvature of the two dimensional euclidean metric 
\begin{align}
d\tilde{s}^2= \ell dw d\bar{w}. \label{stilde}
\end{align}
As noticed in \cite{Klemm}, this action has an interpretation in terms of generalized dilaton theories. 
Following \cite{Grumiller}, it can be shown that the most general solution of the above system is given by
\begin{align}
    \ell(w,\bar w)\equiv \ell(x) &= \frac{1}{4} \left[(x^2 - a^2)^2 + C\right], \\ 
    s &= x - a, \\
    \hat s &= x+a,
\end{align}
where $C$ is an integration constant and the coordinate $x$ is defined by
\begin{align}
2 dx= \ell(x)d(w+\bar w). 
\end{align}
Different types of solutions depend on the sign of $C$. Here, we will concentrate on the case where $C$ is negative and we set
\begin{equation}
 C = -k^4,
\end{equation}
and assume
\begin{equation}
 0 < k^2 < a^2 \label{k}.   
\end{equation}
This choice allows us to obtain solutions with compact $(w, \bar w)$-space. In order to accomplish this we have to choose for $x$ the range 
(granting the positivity of $\ell$)
\begin{equation}
- \sqrt{a^2 -k^2}< x < \sqrt{a^2 -k^2},    
\end{equation}
and to compactify the variable $y = \frac{1}{2i} (w - \bar w) $.
The two dimensional metric $d\tilde s^2$ reads now
\begin{align}
    d\tilde s^2= 4\frac{dx^2}{(x^2 - a^2)^2 -k^4} + \frac 14 \left[(x^2 - a^2)^2 -k^4\right] dy^2 \label{MetricA}. 
\end{align}
At the boundary of the range of $x$, $\ell(x)$ vanishes so that we have to be careful with $y$ in order to avoid singularities. By symmetry, it is sufficient to
investigate what happens at $x\sim \sqrt{a^2 -k^2}$. 
Setting $x= \sqrt{a^2 -k^2} - \xi$, we get for $\xi\sim 0^+$
\begin{equation}
    d \tilde s^2 \approx \frac{4}{(a^2 -k^2)^{\frac{3}{2}}} \left(  (d\sqrt{\xi})^2+ \frac{k^4(a^2-k^2) \xi}{4} dy^2 \right). 
\end{equation}
After introducing the new coordinates
\begin{align}
     \phi &= \frac{k^2 \sqrt{a^2 - k^2}}{2} y \\
     r &= \sqrt{\xi}
\end{align}
we see that the metric is 
\begin{equation}
 d \tilde s^2 \propto dr^2 + r^2 d\phi^2 
\end{equation}
so that, in order to avoid conical singularities, we have to take $\phi$ periodic with period $2 \pi$. With this choice eq.(\ref{MetricA}) defines a smooth compact surface, and applying the Gauss-Bonnet theorem we easily find it is a surface of genus 0.
\newline
Finally we can solve the rotation equation (\ref{1}), that gives
\begin{equation}
    \sigma =\frac{2 \ell}{ G \sqrt{h}} dy. 
\end{equation}
Using the coordinates $t,z,x,\phi$ we find the complete solution
\begin{align}
    ds^2 &= - \frac{2 G (\alpha z + \beta)^2}{a^2-x^2} (dt + \sigma)^2 + \frac{2(a^2-x^2)}{G} \left[\frac{dz^2}{(\alpha z +\beta)^2} + \frac{dx^2}{\ell(x)} + \frac{4 \ell(x)}{k^4(a^2-k^2)} d\phi^2 \right], \label{MetricR} \\
    \tau_1 &= \sqrt{\frac{g_0 g_1}{g_2 g_3}} \left(\frac{2 i \alpha+ a-x}{a+x} \right), \qquad \tau_2=\sqrt {\frac {g_0g_2}{g_1g_3}}, \qquad \tau_3=\sqrt {\frac {g_0g_3}{g_1g_2}}, \\
    \sigma &= \frac{(x^2 - a^2)^2 -k^4}{G k^2 \sqrt{a^2 - k^2}(\alpha z + \beta)} d\phi, \qquad G = 64 \sqrt{g_0 g_1 g_2 g_3}, \\
    \ell(x) &=  \frac{1}{4} \left[(x^2 - a^2)^2 -k^4\right], 
\end{align}
with $\alpha$, $\beta$, $a$ and $k$ constant satisfying eq.(\ref{k}), and the gauge fields are
\begin{align}
     G^A &= -\frac{ G (\alpha z + \beta)}{8 g_A (a-x)^2} dt \wedge dx - \frac{G \alpha}{8 g_A (a-x)} dt \wedge dz - \frac{k^2}{ 8  g_A \sqrt{a^2 - k^2} (a-x)^2} dx \wedge d\phi, \label{GA}\\
     G^B &= - \frac{ G (\alpha z + \beta)} { 8 g_B (a+x)^2} dt \wedge dx + \frac{G \alpha}{8 g_B (a+x)} dt \wedge dz - \frac{k^2}{ 8 g_B \sqrt{a^2 - k^2} (a+x)^2} dx \wedge d\phi, \label{GB}
\end{align}
with duals $\star G_I=Im\mathcal N_{IJ} \star G^J$
\begin{align}
    \star G_A &= -\frac{16 g_A \alpha (\alpha z + \beta)}{(a+x)(a^2-x^2)} dt \wedge dx + 4 g_A  dt \wedge dz - \frac{16 g_A \alpha k^2}{G \sqrt{a^2 - k^2} (a+x)} \frac {dx \wedge d\phi}{(a^2-x^2)},\label{sGA} \\
    \star G_B &= \frac{16g_B \alpha (\alpha z + \beta)}{(4\alpha^2+(a-x)^2)(a+x)} dt \wedge dx +  \frac{4g_B(a-x)^2}{4\alpha^2+(a-x)^2} dt \wedge dz \cr
    &\quad+\frac{16g_B\alpha k^2}{G \sqrt{a^2 - k^2}(4\alpha^2+(x-a)^2)(a+x)} dx \wedge d\phi,\label{sGB}
\end{align}
where $A=0,1$ and $B=2,3$.
\newline
In Appendix it is shown that the metric (\ref{MetricR}) is free of singularities. It has the structure of a fibration over a two dimensional space, with coordinates $x,\phi$, of metric
\begin{align}
 d\Sigma^2=\frac{2(a^2-x^2)}{G} \ d\tilde s^2=\frac{2(a^2-x^2)}{G} \left[\frac{dx^2}{\ell(x)} + \frac{4 \ell(x)}{k^4(a^2-k^2)} d\phi^2 \right].
\end{align}
We know that $d\tilde s^2$ corresponds to a compact surface of genus zero, and since the conformal factor $2(a^2-x^2)/G$ is regular and strictly positive for $x^2\leq a^2-k^2$, the same is true for $d\Sigma^2$. So, the base of the fibration has the topology
of a sphere $S^2$. The corresponding scalar curvature is
\begin{align}
 R_{\Sigma}=\frac G{4(a^2-x^2)} \left[ 3a^2-9x^2 -k^4\frac {a^2+x^2}{(a^2-x^2)^2} \right],
\end{align}
which becomes singular in $x=\pm a$ for $k=0$. This is why we keep $k\neq 0$.\\
As for what concerns the fibres, we see that fixing $x=x_0$, $\phi=\phi_0$, (\ref{MetricR}) reduces to
\begin{align}
 ds^2|_{x_0,\phi_0} &= - \frac{2 G (\alpha z + \beta)^2}{a^2-x_0^2} dt^2 + \frac{2(a^2-x_0^2)}{G} \frac{dz^2}{(\alpha z +\beta)^2},
\end{align}
which is an $AdS_2$ with curvature
\begin{align}
 R_{AdS}=\frac {\alpha^2 G}{a^2-x_0^2}.
\end{align}
So, topologically, for $\alpha\neq 0$ this solution looks to be a deformed Bertotti-Robinson spacetime. However, it has not the same topology but it more resembles the Kerr throat geometry studied in \cite{Bardeen}. To see this let us first notice that,
after redefining the coordinate
\begin{align}
 r=\frac {G\alpha}{k^2} (\alpha z+\beta),
\end{align}
we can rewrite (\ref{MetricR}) as
\begin{align}
 ds^2=\frac {2(a^2-x^2)}{G\alpha^2} \left[ -r^2 dt^2+\frac {dr^2}{r^2} +\alpha^2 \frac {dx^2}{\ell(x)} \right]+\frac {8\ell(x)}{G(a^2-k^2)(a^2-x^2)} \left( d\phi -\frac {\sqrt {a^2-k^2}}\alpha r dt \right)^2.
\end{align}
Then, we introduce a coordinate $\psi$ such that
\begin{align}
 \alpha \frac {dx}{\ell(x)}=d\psi.
\end{align}
A simple calculation shows that 
\begin{align}
 x=\sqrt{a^2-k^2}\ {\rm sn} \left( \frac {\sqrt {a^2+k^2}}{2\alpha} \psi; \sqrt{\frac {a^2-k^2}{a^2+k^2}} \right),
\end{align}
where sn$(z,k)$ is the Jacobi elliptic function defined for $ k^2<1$, by
\begin{align}
 z=\int_0^{{\rm sn}(z;k)} \frac {d\theta}{\sqrt {1-k^2 \sin^2\theta}}.
\end{align}
As usual, we will omit the elliptic modulus, writing for brevity
\begin{align}
 x=\sqrt{a^2-k^2}\ {\rm sn} (u), \qquad\ u= \frac {\sqrt {a^2+k^2}}{2\alpha} \psi.
\end{align}
Therefore, 
\begin{align}
 ds^2=&2\frac {a^2-(a^2-k^2){\rm sn}^2(u)}{G\alpha^2} \left[ -r^2 dt^2+\frac {dr^2}{r^2} +d\psi^2 \right]\cr
 &+\frac {2{\rm cn}^2(u)}{G}\left[ 1+\frac {k^2}{a^2-(a^2-k^2){\rm sn}^2(u)}\right] \left( d\phi -\frac {\sqrt {a^2-k^2}}\alpha r dt \right)^2.\label{metrica sn}
\end{align}
Finally, following \cite{Bardeen}, see also \cite{Chow-4}, we also introduce the new coordinates $\tau, y$ and $\varphi$ by means of the relations
\begin{align}
 r&=y+\sqrt{1+y^2}\ \cos \tau, \\
 t&=\frac {\sqrt {1+y^2}}r \sin \tau, \\
 \phi&=\varphi-\frac {\sqrt {a^2-k^2}}\alpha \log \left| \frac {\cos\tau+y\sin \tau}{1+\sqrt {1+y^2}\ \sin \tau} \right|,
\end{align}
so that 
\begin{align}
 ds^2=&2\frac {a^2-(a^2-k^2){\rm sn}^2(u)}{G\alpha^2} \left[ -(1+y^2) d\tau^2+\frac {dy^2}{1+y^2} +d\psi^2 \right]\cr &+\frac {2{\rm cn}^2(u)}{G}\left[ 1+\frac {k^2}{a^2-(a^2-k^2){\rm sn}^2(u)}\right] \left( d\varphi -y d\tau \right)^2.\label{metrica finale}
\end{align}
We recall that here $u= \frac {\sqrt {a^2+k^2}}{2\alpha} \psi$, and the range of coordinates is
\begin{align}
 -\infty &<\tau<\infty,\\
 -\infty &<y<\infty,\\
 -\frac {2\alpha}{\sqrt{a^2+k^2}} K(\sqrt{\frac {a^2-k^2}{a^2+k^2}}) &\leq \psi \leq \frac {2\alpha}{\sqrt{a^2+k^2}} K(\sqrt{\frac {a^2-k^2}{a^2+k^2}}), \\
 0 &\leq \varphi < 2\pi,
\end{align}
where $K(k)$ is the complete elliptic integral of the first kind. Exactly in the same way as in \cite{Bardeen}, we get that these coordinates cover the whole spacetime, which is geodetically complete. $\tau$ is the global time.
The topology of this solution is not exactly the one of
a Bertotti-Robinson spacetime, but is the same as the Kerr-Newman throat (4.2) of \cite{Bardeen}. This allows us to the following conjecture: our solution is the near horizon limit of a rotating, possibly non BPS, extremal $AdS$ black hole carrying both 
electric and magnetic fields and coupled to a complex scalar field that is non constant at the horizon. Indeed, it is well known that this supergravity theory admits vacua with asymptotic $AdS_4$ regions. 
For the same reasons we have discussed for the static case, the interpolating solutions are now expected to be non BPS. Indeed, this time the Killing spinors must satisfy two projection conditions, since they must satisfy
the conditions $\Gamma^2\Gamma^3\epsilon=\Gamma^2 \Gamma_5\epsilon=i\epsilon$, as one can easily check by explicitly solving equations (3.9) and the subsequent one in \cite{Cacciatori}. The double projection, w.r.t. the static case, reduces the number 
of symmetries down to 1/4 BPS. This can also be checked by proving that, indeed, the stationary solution does not solve equations (3.138) - (3.140) of \cite{Klemm2}.\\
Since in absence of hypermultiplets and nonabelian gaugings there cannot be $1/8$ BPS solutions, see \cite{Cacciatori}, the complete flux solutions are expected
to be non BPS.
Finding them is not trivial. Strategies have been proposed, for example, in \cite{Klemm}, \cite{Hristov}, and \cite{Klemm2}, see also \cite{Klemm:2012yg}, \cite{Klemm:2012vm}, \cite{Klemm:2016wng}. 
However, all these cases start from 1/2 BPS near horizon 
backgrounds and the same strategy does not work for less supersymmetric starting near horizon solutions. Of course, the area of the horizon of the alleged rotating black hole must coincide with the area of the spheroids of the near horizon solution, which 
is 
\begin{align}
 A_{R_H}=\frac {16\pi}{G}.
\end{align}

For $\alpha=0$ the fibres become two dimensional flat Minkowski spacetimes. In this case $\beta\neq0$ and can be rescaled to 1 by a coordinate redefinition. The metric then takes the form
\begin{align}
 ds^2=&2\frac {a^2-(a^2-k^2){\rm sn}^2(u)}{G\alpha^2} \left[ -dt^2+dz^2+d\psi^2 \right]\cr &+\frac {2{\rm cn}^2(u)}{G}\left[ 1+\frac {k^2}{a^2-(a^2-k^2){\rm sn}^2(u)}\right] \left( d\phi -\sqrt {a^2-k^2} dt \right)^2. \label{metric-flat}
\end{align}

\subsection{Symmetries and conservation laws}
If $T_{\mu\nu}$ is the energy-momentum tensor, we can construct the tensor density $\mathcal T_{\mu\nu}$, which satisfies 
\begin{align}
\partial_t \mathcal T^t_{\ \ \lambda}=-\partial_j \mathcal T^j_{\ \ \lambda} +\frac 12 \mathcal T^{\mu\nu} \partial_\lambda g_{\mu\nu}, 
\end{align}
where repeated latin index run from 1 to 3, while the greek ones from 0 to 3. If $S$ is a compact 3-dimensional spatial region with boundary $\partial S$ with normal directions $n_j$ and area element $d\hat \sigma$, we can rewrite it in integral form
\begin{align}
 \frac {d}{dt} \int_S d^3 x\ \mathcal T^t_{\ \ \lambda}=-\oint_{\partial S} d\hat \sigma\ \mathcal T^j_{\ \ \lambda} n_j +\frac 12 \int_S d^3x\  \mathcal T^{\mu\nu} \partial_\lambda g_{\mu\nu}.
\end{align}
For $\lambda=t,y$ the last term vanishes identically and the l.h.s. of the above equation individuate conserved quantities.\footnote{Indeed, $\partial_t$ and $\partial_y$ are Killing vector fields.} \\
The energy-momentum tensor reads
\begin{align}
T_{\mu \nu} =& -(Im \mathcal N)_{IJ} G^I_{\mu\sigma} G_\nu^{J\sigma} + 2 g_{\alpha\bar \beta}\partial_{(\mu} z^\alpha \partial_{\nu)} \bar z^{\bar \beta} \cr &+ g_{\mu \nu} \left[   \frac 14 (Im \mathcal N)_{IJ} G^I_{\sigma\rho} G^{J\sigma\rho}
-g_{\alpha\bar \beta}\partial_\sigma z^\alpha \partial^\sigma \bar z^{\bar \beta} -V \right],
\end{align}
so, the full tensor density $\mathcal T=\sqrt{\gamma} T$ is\footnote{in coordinates $t,z,x,\phi$; here $\gamma$ is minus the determinant of the spacetime metric}
\begin{align}
\mathcal T^\mu_{\ \ \nu}= \begin{pmatrix}
  \mathcal T^t_{\ \ t}  &0 & 0& 0 \\
 0& \mathcal T^z_{\ \ z}  &0 &0 \\
0 & 0& \mathcal T^x_{\ \ x} &0 \\
\mathcal T^\phi_{\ \ t} &0&0& \mathcal T^\phi_{\ \ \phi} \\
\end{pmatrix},
\end{align}
with 
\begin{align}
 \mathcal T^t_{\ \ t}=&\mathcal T^z_{\ \ z}= \frac{2 \ell(x^2 + \alpha^2)-(a^2-x^2)^3}{(a^2-x^2)^2}, \\
 \mathcal T^x_{\ \ x}=& - \frac{2 \ell(x^2 + \alpha^2)}{(a^2-x^2)^2} + 2 (x^2 + \alpha^2), \\
 \mathcal T^\phi_{\ \ \phi}=& - \frac{2 \ell(x^2 + 2a^2 +3\alpha^2)}{(a^2-x^2)^2} + 2 (x^2 + \alpha^2), \\
 \mathcal T^\phi_{\ \ t}=& -\frac {G(\alpha z+\beta)k^2\sqrt{a^2-k^2} (a^2+x^2+2\alpha^2)}{(a^2-x^2)^2}.
\end{align}
We then see that for $\lambda=\phi$ the conservation equation is just an identity that gives just zero. The only interested conserved quantity is associated to $\mathcal T^t_{\ \ t}$. For 
\begin{align}
 S=\{(z,x,\phi)| z_1\leq z\leq z_2,\ -\sqrt{a^2-k^2} \leq x \leq \sqrt{a^2-k^2},\ 0\leq \phi\leq 2\pi \},
\end{align}
we get 
\begin{align}
\frac {\Theta_S}{2\pi \Delta z} &:=\frac 1{2\pi \Delta z} \int_S d^3x\  \mathcal T^t_{\ \ t}    \cr
&= \left(\alpha^2(1-\frac {k^2}{2a^2}) -a^2 -\frac{3}{2}k^2\right)\sqrt{a^2-k^2} + \frac{a^2-\alpha^2}{4a^3}k^4 \log{\frac{a + \sqrt{a^2-k^2}}{a-\sqrt{a^2-k^2}}},
\end{align}
where $\Delta z=z_2-z_1$. This energy is conserved in any compact region, in a stationary way, since there is a constant flow of energy in the $\phi$ direction, due to the $\mathcal T^\phi_{\ \ t}$ term.\\
Let us give a better look to the Killing vector fields. Looking at (\ref{metrica sn}) we see that beyond translation in $t$ and $\phi$, the metric is invariant under the rescaling 
\begin{align}
 r\mapsto \lambda r, \qquad\ t\mapsto \frac t\lambda.
\end{align}
To this symmetry it correspond the Killing vector $r\partial_r-t\partial_t$. From (\ref{metrica finale}) we see that another invariance is under translation of the global time $\tau$, to which it corresponds the Killing vector field 
$\zeta_2=\partial_\tau$. All Killing vectors are therefore
\begin{align}
 \zeta_0&=\partial_t, \\
 \zeta_1&=r\partial_r-t\partial_t, \\
 \zeta_2&=\frac 12 \left( \frac 1{r^2}+1+t^2\right)\partial_t-tr\partial_r-\frac {\sqrt {a^2-k^2}}{\alpha r} \partial_\phi,\\
 \zeta_3&=\partial_\phi.
\end{align}
They generate an $SL(2,\mathbb R)\times U(1)$ isometry group.\\
When $\alpha=0$, the Killing vectors of (\ref{metric-flat}) are
\begin{align}
  \zeta_0&=\partial_t, \\
 \zeta_1&=\partial_z, \\
 \zeta_2&=z\partial_t+t\partial_z,\\
 \zeta_3&=\partial_\phi,
\end{align}
and generate the isometry group $(O(1,1)\ltimes \mathbb R)\times \mathbb R$.

\subsection{Electromagnetic properties}
Consider now (\ref{GA}) and (\ref{GB}). They have magnetic fluxes
\begin{equation}
    P^I = \frac{1}{4 \pi} \int_\Sigma G^I = -\frac{1}{8 g_I}, \label{magnetic-fluxes}
\end{equation}
which do not correspond to magnetic charges, since $\Sigma$ do not bound any singular region. Analogously, from (\ref{sGA}) and (\ref{sGB}) we can compute the electric fluxes, which again result to vanish, despite in this case
the electric fields are not zero. Indeed, the fields $\mathcal G_I=Re (\mathcal N_{IJ}) G^J-\star G_I$ are
\begin{align}
\mathcal G_A=&-\frac{32 g_A \alpha (\alpha z + \beta)x}{(a^2-x^2)^2} dt \wedge dx - 4 g_A\left(1+\frac {4\alpha^2}{a^2-x^2}\right)  dt \wedge dz \cr
&- \frac{32 g_A \alpha k^2 x}{G \sqrt{a^2 - k^2}} \frac {dx \wedge d\phi}{(a^2-x^2)^2},\\
\mathcal G_B=&  -4g_B dt \wedge dz .
\end{align}
Indeed, the solution being regular everywhere, included the fieldstrengths and their duals, there is not any natural candidate as source of charges (apart at most at infinite in $z$).
The absence of sources of charges is compatible with the interpretation of the solution as a near horizon limit. Of course, the fluxes in (\ref{magnetic-fluxes})
are expected to represent the magnetic 
charges of the alleged rotating black hole having our solution as near horizon limit.  We want to notice that the non vanishing electric field has the peculiarity to have positive flux through every section of the spheroid with $x>0$ and negative 
where $x<0$ (recall that $g_A<0$), and they perfectly compensate globally, exactly as it happens for an axial dipole placed along the axis defining the angle $\phi$.

\section{Conclusions and perspectives}\label{Conclusions}
We have found new interesting supersymmetric backgrounds in $\mathcal N=2$, $D=4$ gauged supergravity coupled to vector multiplets for the STU model with prepotential $F(X^0,\ldots,X^3)=-2i\sqrt{X^0X^1X^2X^3}$. These are everywhere regular
solutions with one nonconstant complex scalar field and both magnetic and electric fields with vanishing fluxes. A parameter $\alpha$ can be set to zero to switch off the electric fields. 
In this limit the topology of the solutions is the one of a fibration $\mathbb R^{1,1}$ over a surface of genus $0$, while
for $\alpha\neq 0$ the solutions have the same topology as the Ker-Newman throat of \cite{Bardeen}: a fibration of $AdS_2$ fibres on the base of conformal spheres with non constant scalar curvature. These are stationary solutions, with a rotation parameter 
along a Killing spacelike direction on the conformal spheres. 
It is interesting to notice that when $\alpha$ is switched on, the total electric flux vanishes, but it is positive everywhere in the north semi spheroid ($x>0$) end negative in the south semi spheroid. It resembles a dipolar electric field.

Notice that the metric in our solution belongs to the general class conjectured in \cite{Gnecchi}, formula (2.1), and there checked for all known rotating black holes in matter-coupled N = 2 gauged supergravity in four dimensions. This general form is
\begin{align}
    ds^2 = - f(z,x) (dt + \omega d\phi)^2 + \frac{1}{f(z,x)}\left[ v(z,x) \left( \frac{dz^2}{Q(z)} + \frac{dx^2}{P(x)} \right) +P(x) Q(z) d\phi^2 \right] \label{genForm}
\end{align}
where $P(x)$ and $Q(z)$ are generic polynomials of fourth-degree. Our (\ref{MetricR}) can be recast in the form (\ref{genForm}) after the identifications
\begin{align}
    P(x)=& \frac{4 c}{k^2 \sqrt{a^2-k^2}} \ell(x), \qquad v(z)= \frac{16 c}{k^2 \sqrt{a^2-k^2}} \left(\alpha z +\beta \right)^2\\
    Q(z)=& \frac{v(z)}{4}, \qquad f(z,x)= \frac{2 G}{a^2-x^2} \left(\alpha z+ \beta \right)^2,  \qquad \omega d\phi = c \, \sigma , \\
    \sigma =& \frac{8}{G} \frac{1}{v(z)} P(x) \left( \alpha z + \beta \right) d\phi, 
\end{align}
where $c$ is a generic constant, which can be fixed choosing the normalization of the $\phi$ coordinate. It is worth to mention that at the end of \cite{Gnecchi}, after analysing all known solutions, they improve their ansatz, proposing the following specifications
\begin{align}
    v=& Q-P, \qquad f = v \exp{(2U)}, \qquad \exp{(-2U)} = e^{-\mathcal K}, \\
    \omega= &- \frac{1}{v} \left[ Q(c_0 +c_1 x+ c_2 x^2) + P(d_0+d_1 z+d_2 z^2)\right].
    \label{AnsatzG}
\end{align}
However, our solution does not fit into these specifications, which, therefore, are not general.
In particular, our solution does not satisfies the first line of these relations. 

It is interesting to notice that these solutions have the same topology of the near horizon geometry of extremal black holes. We indeed conjecture that these solutions are the near horizon limit of some rotating extremal black hole configuration
coupled to electric and magnetic fields. We also mention that the solutions presented here do not admit limits approaching other known solutions. This seems to be related to the fact that the signs of the couplings $g_I$ are responsible of the fact that the 
potential $V$ is positive near the horizon while it is expected to be negative at the $AdS_4$ region of the interpolating solution. This justifies the fact that we get a spherical topology at the horizon, while all solutions found, for example, in \cite{Klemm} have 
hyperbolic topology. This is not surprising since it happens already for static solutions (easier to be found), see section 3.2 in \cite{Cacciatori2}. Also, it is worth to mention that AdS black holes are related to the existence of solitons, solutions that have 
no smooth limit when the coupling constant goes to zero, see for example \cite{Romans}.

It would be interesting to understand if they carry thermodynamical properties. Another possible generalization of the present
paper is the following. In order to solve the equations we assumed $\partial^2_z h=0$, equivalent to the ansatz (\ref{s}), (\ref{s1}). However, equations (\ref{8}) and (\ref{9}) require a weaker assumption, which is 
\begin{align}
 & \partial_z \left( \sqrt{h} \partial^2_z  \sqrt{h} \right)=0, \\
 & \eta = \frac 1{16}\frac{ s( w, \bar w)}{\sqrt {h(z)}}, \qquad  \hat{\eta} = \frac 1{16} \frac{ \hat s(w , \bar w)}{\sqrt {h(z)} }.
\end{align}
Rewriting the first as 
\begin{align}
 \partial^2_z  \sqrt{h}=\frac {\xi}{\sqrt h}
\end{align}
with $\xi$ a constant, we are led to the system
\begin{align}
 \frac {\partial_w\partial_{\bar w} S}{\ell} &=\frac {\xi}4 S+\frac 18 S\hat S(S+\hat S), \\
 \frac {\partial_w\partial_{\bar w} \hat S}{\ell} &=\frac {\xi}4 \hat S+\frac 18 S\hat S(S+\hat S),\\
 \frac {\partial_w\partial_{\bar w} \log \ell}{\ell}&= -\frac \xi2 +\frac 18 (S+\hat S)^2+\frac 14 S\hat S.
\end{align}
This system of equations is much more difficult to solve than the one considered in the present paper. It would be interesting to find nontrivial solutions of this system and to study the corresponding spacetime solutions. These and other questions will be
considered in a forthcoming paper.

\section*{Acknowledgments}
We thank Silke Klemm for helpful discussions. We also thank Alessio Marrani for a kind discussion about reference \cite{Bellucci}, and the anonymous referee for relevant comments.

\begin{appendix}

\section{Curvature}
A vierbein for the metric (\ref{MetricR}) is 
\begin{align}
 E^0& =\sqrt{\frac {2G}{a^2-x^2}}\ (\alpha z+\beta)(dt+\sigma),  \qquad E^1=\sqrt {\frac{2(a^2-x^2)}{G}}\ \frac {dz}{\alpha z+\beta}, \\
 E^2&=\sqrt {\frac{2(a^2-x^2)}{G}}\ \frac {dx}{\sqrt {\ell(x)}}, \qquad\qquad\   E^2=\frac 2{k^2}\sqrt {\frac{2\ell(x)(a^2-x^2)}{G(a^2-k^2)}}\  d\phi.
\end{align}
The connection 1-forms, defined by 
\begin{align}
 dE^a+\omega^a_{\ \ b} \wedge E^b=0, \qquad\ \omega_{ab}=-\omega_{ba},
\end{align}
where the indices $a,b$ run from 0 to 3, and are raised and lowered with the flat Minkowski metric. These are
\begin{align}
\omega^0_{\ 1}&=\alpha \sqrt {\frac {G}{2(a^2-x^2)}}\ E^0 -2 \frac {\alpha \sqrt {\ell(x)}}{G} \left( \frac {G}{2(a^2-x^2)} \right)^{\frac 32}\ E^3,\\
\omega^0_{\ 2}&=\frac {x}{a^2-x^2} \sqrt {\frac {G\ell(x)}{2(a^2-x^2)}}\ E^0 -x\sqrt {\frac {G}{2(a^2-x^2)}}\  E^3,\\
\omega^0_{\ 3}&=2 \frac {\alpha \sqrt {\ell(x)}}{G} \left( \frac {G}{2(a^2-x^2)} \right)^{\frac 32}\ E^1 +x\sqrt {\frac {G}{2(a^2-x^2)}}\  E^2,\\
\omega^1_{\ 2}&=-\frac {x}{a^2-x^2} \sqrt {\frac {G\ell(x)}{2(a^2-x^2)}}\ E^1, \\
\omega^1_{\ 3}&=-2 \frac {\alpha \sqrt {\ell(x)}}{G} \left( \frac {G}{2(a^2-x^2)} \right)^{\frac 32}\ E^0, \\
\omega^2_{\ 3}&= -x\sqrt {\frac {G}{2(a^2-x^2)}}\  E^0 +\sqrt {\frac {G}{2\ell(x)(a^2-x^2)}} \frac x{a^2-x^2}\left( \ell(x) +\frac {a^2-x^2}2 \right) E^3.
\end{align}
The curvature 2-form, defined by 
\begin{align}
 \Omega^a_{\ b} \equiv \frac 12 \Omega^a_{\ bcd} E^c\wedge E^d=d\omega^a_{\ b}+\omega^a_{\ c}\wedge \omega^c_{\ b},
\end{align}
has relevant non vanishing terms 
\begin{align}
 \Omega^0_{\ 101}&=\frac {b^2}{h}\left( \frac {\ell(x)(x^2-\alpha^2)}{(a^2-x^2)^2}  -\alpha^2\right); \qquad\ \Omega^0_{\ 102}=-3\alpha \frac {b^2}{h} \frac {x\sqrt {\ell(x)}}{a^2-x^2}; \\
 \Omega^0_{\ 113}&=\frac {b^2}{h} \frac {x^2-2\alpha^2}{a^2-x^2} \sqrt {\ell(x)}; \qquad\ \Omega^0_{\ 123}=-\frac {b^2}{h} \alpha x \left( 1+\frac {2\ell(x)}{(a^2-x^2)^2} \right); \\
 \Omega^0_{\ 202}&=-\frac {b^2}{h} \left( \ell(x) \frac {a^2+3x^2}{(a^2-x^2)^2}+\frac {x^2}2 \right); \qquad\ \Omega^0_{\ 213}=-\frac {b^2}{h} \alpha x \left( \frac 12 +\frac {4\ell(x)}{(a^2-x^2)^2} \right);\\ 
 \Omega^0_{\ 223}&=-\frac {b^2}{h} \frac {a^2+2x^2}{(a^2-x^2)} \sqrt {\ell(x)}; \qquad\ \Omega^0_{\ 312}=\frac {b^2}{h} \alpha x \left( \frac 12 -\frac {2\ell(x)}{(a^2-x^2)^2} \right); \\
 \Omega^0_{\ 303}&=\frac {b^2}{h} \left( \ell(x) \frac {x^2-\alpha^2}{(a^2-x^2)^2} -\frac {x^2}2\right); \qquad\ \Omega^1_{\ 212}=\frac {b^2}{h} \left( \ell(x) \frac {x^2+a^2}{(a^2-x^2)^2} -\frac {x^2}2\right);\\
 \Omega^1_{\ 313}&=\frac {b^2}{h} \left( \ell(x) \frac {3\alpha^2-x^2}{(a^2-x^2)^2} -\frac {x^2}2\right); \qquad\ \Omega^1_{\ 323}=\frac {b^2}{h} 3\alpha x\frac {\sqrt {\ell(x)}}{a^2-x^2}; \\
 \Omega^2_{\ 323}&=\frac {b^2}{h} \left( \ell(x) \frac {a^2+x^2}{(a^2-x^2)^2} +\frac {a^2+x^2}2 \right),
\end{align}
where
\begin{align}
 \frac {b^2}h=\frac {G}{2(a^2-x^2)}.
\end{align}
Looking at this expression it is evident that the solution is completely regular if $k\neq 0$, while for $k=0$ it becomes singular at $x=\pm a$. \\
In particular, the Ricci scalar is
\begin{align}
 R=\frac {G}{2(a^2-x^2)} \left[ a^2-3x^2 -2\alpha^2 +2\ell(x) \frac {a^2+\alpha^2}{(a^2-x^2)^2} \right].
\end{align}
 
\end{appendix}



\begin{thebibliography}{99}

\bibitem{Hartnoll}
S. A. Hartnoll, 
``Lectures on holographic methods for condensed matter physics,'' 
Class. Quant. Grav. 26 (2009) 224002

\bibitem{Iizuka}
N. Iizuka, N. Kundu, P. Narayan and S. P. Trivedi, ``Holographic Fermi and non-Fermi liquids with transitions in dilaton gravity,'' JHEP 1201 (2012) 094

\bibitem{Hartnoll2}
S. A. Hartnoll, C. P. Herzog and G. T. Horowitz, ``Building a holographic superconductor,'' Phys. Rev. Lett. 101 (2008) 031601

\bibitem{Guica}
M. Guica, T. Hartman, W. Song and A. Strominger, ``The Kerr/CFT correspondence,'' Phys. Rev. D 80 (2009) 124008

\bibitem{Benini}
  F.~Benini, K.~Hristov and A.~Zaffaroni,
  ``Black hole microstates in AdS$_4$ from supersymmetric localization,''
  JHEP 1605 (2016) 054

\bibitem{Bellucci}
S.~Bellucci, S.~Ferrara, A.~Marrani and A.~Yeranyan,
``d=4 Black Hole Attractors in N=2 Supergravity with Fayet-Iliopoulos Terms,''
Phys. Rev. D \textbf{77} (2008), 085027

\bibitem{Cacciatori2}
S.~L.~Cacciatori and D.~Klemm,
  ``Supersymmetric AdS(4) black holes and attractors,''
  JHEP {\bf 1001} (2010) 085

\bibitem{Kachru}  
S. Kachru, R. Kallosh and M. Shmakova, ``Generalized attractor points in gauged supergravity,'' Phys. Rev. D 84 (2011) 046003

\bibitem{Hristov}
K.~Hristov, S.~Katmadas and C.~Toldo,
  ``Matter-coupled supersymmetric Kerr-Newman-AdS$_4$ black holes,''
  Phys.\ Rev.\ D {\bf 100} (2019) no.6,  066016

\bibitem{Gnecchi}
A.~Gnecchi, K.~Hristov, D.~Klemm, C.~Toldo and O.~Vaughan,
  ``Rotating black holes in 4d gauged supergravity,''
  JHEP {\bf 1401} (2014) 127

\bibitem{Cacciatori}
S.~L.~Cacciatori, D.~Klemm, D.~S.~Mansi and E.~Zorzan,
  ``All timelike supersymmetric solutions of N=2, D=4 gauged supergravity coupled to abelian vector multiplets,''
  JHEP {\bf 0805} (2008) 097

\bibitem{Daniele}
N.~Daniele, F.~Faedo, D.~Klemm and P.~F.~Ram\`\i rez,
  ``Rotating black holes in the FI-gauged $N=2$, $D=4$ $\overline{\mathbb{C}\text{P}}^n$ model,''
  JHEP {\bf 1903} (2019) 151

\bibitem{Klemm}
D.~Klemm,
  ``Rotating BPS black holes in matter-coupled $AdS_4$ supergravity,''
  JHEP {\bf 1107} (2011) 019

\bibitem{Chow-1}
D.~D.~K.~Chow and G.~Comp\'ere,
``Dyonic AdS black holes in maximal gauged supergravity,''
Phys. Rev. D \textbf{89} (2014) no.6, 065003

\bibitem{Chow-2}
D.~D.~Chow,
``Single-charge rotating black holes in four-dimensional gauged supergravity,''
Class. Quant. Grav. \textbf{28} (2011), 032001

\bibitem{Chow-3}
D.~D.~Chow,
``Two-charge rotating black holes in four-dimensional gauged supergravity,''
Class. Quant. Grav. \textbf{28} (2011), 175004

\bibitem{Chong}
Z.~W.~Chong, M.~Cvetic, H.~Lu and C.~Pope,
``Charged rotating black holes in four-dimensional gauged and ungauged supergravities,''
Nucl. Phys. B \textbf{717} (2005), 246-271

\bibitem{Hristov-1}
K.~Hristov, S.~Katmadas and C.~Toldo,
``Rotating attractors and BPS black holes in $AdS_4$,''
JHEP \textbf{01} (2019), 199

\bibitem{Ferrara:1997yr}
  S.~Ferrara,
  ``Bertotti-Robinson geometry and supersymmetry,''
  hep-th/9701163.

\bibitem{Bardeen}
J.~Bardeen and G.~T.~Horowitz,
``Extreme Kerr throat geometry: A vacuum analog of $AdS_2\times S^2$,''
Phys.\ Rev.\ D {\bf 60} (1999) 104030

\bibitem{Chow-4}
D.~D.~Chow, M.~Cvetic, H.~Lu and C.~Pope,
``Extremal Black Hole/CFT Correspondence in (Gauged) Supergravities,''
Phys. Rev. D \textbf{79} (2009), 084018

\bibitem{Clement:2001gia}
  G.~Clement and D.~Gal'tsov,
  ``Bertotti-Robinson type solutions to dilaton - axion gravity,''
  Phys.\ Rev.\ D {\bf 63} (2001) 124011

\bibitem{Bouchareb:2014fxa}
  A.~Bouchareb, C.~M.~Chen, G.~ClÃ©ment and D.~Gal'tsov,
  ``Bertotti-Robinson and soliton string solutions of $D=5$ minimal supergravity,''
  Phys.\ Rev.\ D {\bf 90} (2014) no.2,  024047

\bibitem{Clement:2001ny}
  G.~Clement and D.~Gal'tsov,
  ``Conformal mechanics on rotating Bertotti-Robinson space-time,''
  Nucl.\ Phys.\ B {\bf 619} (2001) 741

\bibitem{Kunduri:2007vf}
  H.~K.~Kunduri, J.~Lucietti and H.~S.~Reall,
  ``Near-horizon symmetries of extremal black holes,''
  Class.\ Quant.\ Grav.\  {\bf 24} (2007) 4169

\bibitem{DallAgata:2010ejj}
G.~Dall'Agata and A.~Gnecchi,
``Flow equations and attractors for black holes in N = 2 U(1) gauged supergravity,''
JHEP \textbf{03} (2011), 037

\bibitem{Klemm:2012yg}
D.~Klemm and O.~Vaughan,
``Nonextremal black holes in gauged supergravity and the real formulation of special geometry,''
JHEP \textbf{01} (2013), 053

\bibitem{Klemm:2012vm}
D.~Klemm and O.~Vaughan,
``Nonextremal black holes in gauged supergravity and the real formulation of special geometry II,''
Class. Quant. Grav. \textbf{30} (2013), 065003

\bibitem{Klemm:2016wng}
D.~Klemm, N.~Petri and M.~Rabbiosi,
``Symplectically invariant flow equations for $N = 2$, $D = 4$ gauged supergravity with hypermultiplets,''
JHEP \textbf{04} (2016), 008

\bibitem{Grumiller}
D.~Grumiller, W.~Kummer and D.~V.~Vassilevich,
 ``Dilaton gravity in two-dimensions,''
 Phys.\ Rept.\  {\bf 369} (2002) 327

\bibitem{Klemm2}
D.~Klemm and E.~Zorzan,
``The timelike half-supersymmetric backgrounds of N=2, D=4 supergravity with Fayet-Iliopoulos gauging,''
Phys. Rev. D \textbf{82} (2010), 045012

\bibitem{Romans}
L. J. Romans, ``Supersymmetric, cold and lukewarm black holes in cosmological
Einstein-Maxwell theory,'' Nucl. Phys. B 383 (1992) 395

\end{thebibliography}
\end{document}